\documentclass[aps,prl,twocolumn,superscriptaddress]{revtex4-2}
\usepackage{amsmath}
\usepackage{color}
\usepackage{bbm}
\usepackage{amssymb}
\usepackage{epsfig}
\usepackage{multirow}
\usepackage{amsbsy}
\usepackage{array}
\usepackage{diagbox}
\usepackage{bm}
\usepackage{extarrows}
\usepackage{graphicx}
\usepackage{appendix}
\usepackage{txfonts}
\usepackage[colorlinks=true,linkcolor=blue,citecolor=blue,urlcolor=blue,bookmarks=true]{hyperref}

\begin{document}
	\title{Fractionalized Prethermalization in a Driven Quantum Spin Liquid}
	
	\author{Hui-Ke Jin}
	\affiliation{Technical University of Munich, TUM School of Natural Sciences, Physics Department, 85748 Garching, Germany}
	
	\author{Johannes Knolle}
	\affiliation{Technical University of Munich, TUM School of Natural Sciences, Physics Department, 85748 Garching, Germany}
	\affiliation{Munich Center for Quantum Science and Technology (MCQST), Schellingstr. 4, 80799 M{\"u}nchen, Germany}
	\affiliation{Blackett Laboratory, Imperial College London, London SW7 2AZ, United Kingdom}
	
	\author{Michael Knap}
	\affiliation{Technical University of Munich, TUM School of Natural Sciences, Physics Department, 85748 Garching, Germany}
	\affiliation{Munich Center for Quantum Science and Technology (MCQST), Schellingstr. 4, 80799 M{\"u}nchen, Germany}
	
	\date{\today}
	\begin{abstract}
		Quantum spin liquids subject to a periodic drive can display fascinating non-equilibrium heating behavior because of their emergent fractionalized quasi-particles. Here, we investigate a driven Kitaev honeycomb model and examine the dynamics of emergent Majorana matter and $Z_2$ flux excitations. We uncover a distinct two-step heating profile -- dubbed fractionalized prethermalization -- and a quasi-stationary state with vastly different temperatures for the matter and the flux sectors. We argue that this peculiar prethermalization behavior is a consequence of fractionalization. Furthermore, we discuss an experimentally feasible protocol for preparing a zero-flux initial state of the Kiteav honeycomb model with a low energy density, which can be used to observe fractionalized prethermalization in quantum information processing platforms.
	\end{abstract}
	
	\maketitle
	
	{\textbf{ \em Introduction.---}} Coherent time-periodic modulations have been established over the last years as a versatile tool for engineering new Hamiltonians for sought-after equilibrium phases of matter~\cite{Oka2009, Lindner2011, Goldman2014, Bukov2015, Eckardt2017, Cooper2019, Oka2019, Rudner2020}, as well as for realizing novel dynamical topological phases which do not possess an equilibrium analogue~\cite{Kitagawa2010,Jiang2011,Gomez-Leon2013,Rudner2013,Keyserlingk2016,Potter2016,Roy2016,Po2016,Else2016PRB,Roy2017,Roy2017PRB2,Harper2017,Lindner2017, Esin2018,Zhang2021}. Experimental demonstrations include the manipulation of Dirac cones by circularly polarized light~\cite{Wang2013, Mciver2020} and the realization of topological band structures with ultracold atoms~\cite{Aidelsburger2013, Miyake2013, Jotzu2014, Wintersperger_2020}. Recent work has proposed to realize exotic interacting Floquet phases with intrinsic topological order, characterized by fractionalized excitations, including fractional Chern insulators~\cite{Grushin2014}, quantum spin liquids~\cite{Po2017, Claassen_2017, Fidkowski2019, Fulga2019, Sriram2021, Bostrom2022, Kalinowski2022, Sun_2022, Petiziol_2022} and Floquet fracton codes~\cite{Zhang2022Code}. 
	
	A major challenge for Floquet engineering concerns heating due to the continuous energy absorption from the periodic modulation which necessarily drives the system at some point to a featureless infinite-temperature state. Nonetheless, non-trivial Floquet phases can be protected by either many-body localization in the presence of strong disorder~\cite{bordia2017periodically, Ponte2015, Lazarides2015, Gopalakrishnan2016} or by resorting to a high-frequency modulation, that drives the system into a prethermal regime for an exponentially long time~\cite{Bukov2015PRL, Abanin2015, Canovi2016, Mori2016, Weidinger2017, Abanin2017CMP, Mori2018, Howell2019, Andrea2021, Ye2021}. In generic quantum and classical many-body systems with homogeneous energy absorption, the prethermal regime arises at intermediate time scales leading to a quasi-stationary state described by a low-temperature thermal Gibbs ensemble of an effective Hamiltonian~\cite{Kuhlenkamp2020}. However, for Floquet multi-band systems a situation can arise in which the energy bands are at vastly different temperatures; as for example shown for the partially-filled interacting Thouless pump~\cite{Lindner2017}.
	In general, prethermalization in driven systems with inhomogenous energy absorption stemming from different types of excitations remains largely unexplored. This raises the question whether driven topological phases may exhibit prethermal regimes described by effective Gibbs states or whether novel types of heating dynamics can emerge especially in the presence of fractionalized excitations?
	
	\begin{figure}[t!]
		\centering
		\includegraphics[width=1.\linewidth]{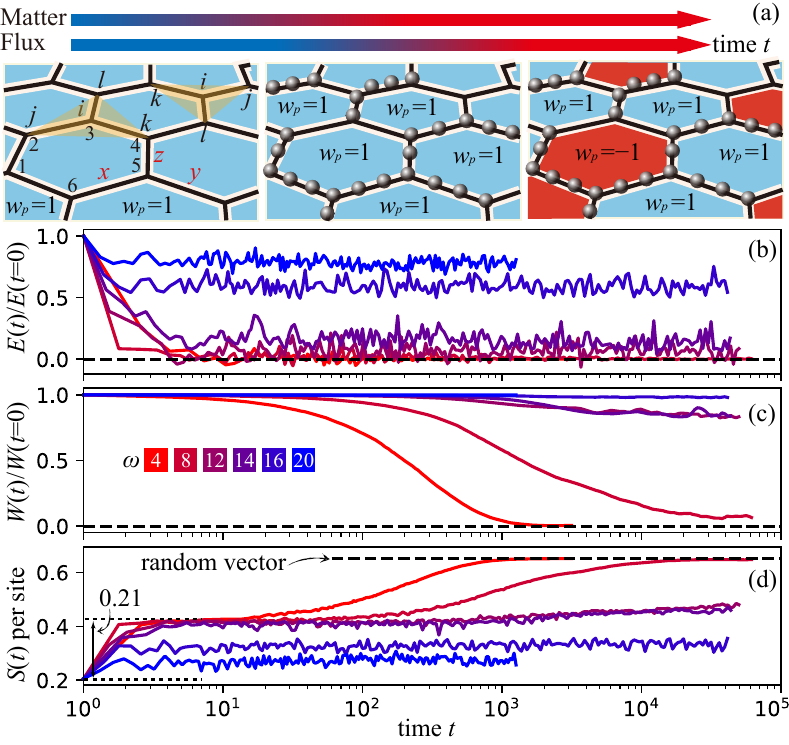}
		\caption{ \textbf{Fractionalized prethermalization.} (a) Schematic heating dynamics of a driven Kitaev honeycomb model with fractionalized matter and flux excitations. 
			{\bf Left:} 
			Ground state characterized by uniform fluxes $w_p=1$ (blue hexagons) and no matter excitations. 
			{\bf Middle:} After a quick relaxation, fluxes remain frozen (blue hexagons) over an exponentially long timescale yet matter fermions (gray spheres) are already thermally activated -- the {\it fractional prethermal} regime.
			{\bf Right:} At late times, fluxes are excited ($w_p=-1$, red hexagons) and the system eventually heats up to a global infinite temperature state. 
			(b, c) For intermediate frequencies, we observe a regime in which the energy $E(t)$ is close to zero indicating a near-infinite temperature state of the fermionic matter, while fluxes $W(t)$ remain close to their ground state. The two fractionalized excitations thus are described by different temperatures, even though the physical spin degree of freedom is driven.
			(d)  The growth of entanglement entropy density shows a multi-stage heating dynamics of the system. The simulations are performed for $V=1.0$ and $J=0.02$.
		}
		\label{fig:fig1}
	\end{figure}

	In this work, we show that the energy absorption of driven fractionalized phases can generically be quite intricate. In particular, we establish that in a periodically driven system with intrinsic topological order, long-lived quasi-steady states can be attained in which the fractionalized excitations are at vastly different temperatures -- a phenomenon we dub {\it fractionalized prethermalization}. 
	To illustrate this unconventional prethermal regime, we consider a periodically driven Kitaev honeycomb model, in which spins fractionalize into emergent matter fermions and  $Z_2$ fluxes. When driving the Kitaev honeycomb model we find situations in which the matter sector heats swiftly while the fluxes remain at low temperatures, realizing a fractionalized prethermalization regime in which the two emergent degree of freedoms are described by vastly different temperatures. We argue that fractionalized prethermalization not only extends the known phenomenology of heating dynamics, but can in turn be used as a tool for diagnosing the presence of fractionalized excitations in 
	quantum simulator platforms.

	{\textbf{ \em Model.---}}We consider the Kitaev honeycomb model~\cite{Kitaev06,hermanns2018physics} as an archetypal, solvable model with topological order:
	\begin{equation}
		H_{\rm K}=-\sum_{a=x,y,z}\sum_{\langle{}ij\rangle_a}K_a\sigma^a_j{}\sigma^a_k,\label{eq:HK}
	\end{equation}
	where $\bm{\sigma}_j=(\sigma^x_j,\sigma^y_j,\sigma^z_j)$ are three Pauli matrices and $\langle{}jk\rangle_a$  ($a=x,y,z$) denotes the $a$-type Ising interactions on an $a$-type bond, see Fig.~\ref{fig:fig1}(a). In our work we concentrate on isotropic interactions $K_x=K_y=K_z=K=1$. There exist commuting plaquette operators on each hexagon $p$, $W_p\equiv\sigma^x_1\sigma^y_2\sigma^z_3\sigma^x_4\sigma^y_5\sigma^z_6$,  with sites labeled as shown in Fig.~\ref{fig:fig1}(a). The plaquette operators $W_p$ commute with the Hamiltonian in Eq.~\eqref{eq:HK}, $[W_p, H_{\rm K}]=0$, which shows that the Hilbert space of $H_{\rm K}$ can be block-diagonalized into orthogonal sectors characterized by the conserved $Z_2$ fluxes $\{w_p=\pm{}1\}$ with $w_p$ the eigenvalue of $W_p$. 
	
	The Kitaev honeycomb model~\eqref{eq:HK} can be solved by introducing the four-Majorana representation~\cite{Kitaev06} $\sigma^{a}_j=ic^{a}_jc^{0}_j$, where $c^a$ ($c^0$) are so-called gauge (itinerant) Majorana fermions. Under this representation, ${H}_{\rm K}$ becomes a quadratic Hamiltonian of itinerant Majorana fermions coupled to a static $Z_2$ gauge field: $H_{\rm K} = -iK\sum_a\sum_{\langle jk \rangle\in a}{u}_{jk} c^0_j{}c^0_k$,
	where $u_{jk} \equiv ic_{j}^{a}c^{a}_{k}$ lives on an $a$-type bond with eigenvalues $u_{jk}=\pm{}1$. Moreover, the plaquette operator $W_p$ can be expressed as a product of $u_{jk}$ around hexagon $p$, i.e., $W_p=\prod_{\langle{}jk\rangle\in{}\partial{}p}u_{jk}$~\cite{Kitaev06}. This representation introduces unphysical states accompanied by gauge redundancy~\cite{Wen02}. The physical Hilbert space can be restored by imposing a local constraint $D_j\equiv c^{x}_jc^{y}_jc^{z}_jc^0_j=1$ at each lattice site $j$. 
	Since all $u_{ij}$'s commute with each other, the eigenstates of $H_{\rm K}$ are obtained by a Bogoliubov-de-Gennes transformation after fixing all gauge fields, for instance, as $u_{ij}=1$. It indicates that in the Kitaev honeycomb model, the spin degrees of freedom are fully fractionalized into the gauge and matter sectors~\cite{baskaran2007exact}. The ground state $|\Psi_{0}\rangle$ is thus a zero-flux state with all $w_p=1$. 
	
	Our goal is to study the dynamics of the Kitaev spin liquid phase with fractionalized gauge and matter excitations under a nonequilibrium drive. We are interested in the generic heating behavior beyond the the fine-tuned point of the pure integrable Kitaev honeycomb model. To this end, we consider the model in Eq.~\eqref{eq:HK} subjected to a periodic modulation at frequency $\omega=2\pi/T$
	\begin{equation}
		H(t)= \begin{cases}
			H_{\rm K}+H_{\rm J}+H_{\rm V} & {\rm for\ } t\in[0,T/2), \\
			H_{\rm K}-H_{\rm J}-H_{\rm V} & {\rm for\ } t\in[T/2,T).
		\end{cases}\label{eq:Ht}
	\end{equation}
	The modulation is generated by two additional terms. First, the Heisenberg interaction $H_{\rm J}=J\sum_{\langle{}jk\rangle}{\bm \sigma}_j\cdot{}{\bm \sigma}_k$ on nearest-neighbor bonds $\langle{}jk\rangle$, which breaks the flux conservation and is expected to heat both the flux and matter sectors. Second, in order to allow for inhomogeneous energy absorption we include the three-spin interaction defined as
	\begin{equation}
		H_{\rm V}=V\sum_{\langle{}jkl\rangle\in{}i}\sigma^x_j{}\sigma^y_k\sigma^z_l,~\label{eq:HV}
	\end{equation}
	where $\langle{}jkl\rangle\in{}i$ denotes the spin triples on the vortices (with center $i$) of the honeycomb lattice, as graphically indicated in Fig.~\ref{fig:fig1}(a). In the Majorana representation, $H_{\rm V}$ can be rewritten as
	$H_{\rm V}=V\sum_{\langle{}jkl\rangle\in{}i}u_{ki}u_{ji}u_{li}c^0_ic^0_jc^0_kc^0_l,$ where site $i$ is the center of triangle $\langle{}jkl\rangle\in{}i$.
	Therefore, $H_{\rm V}$ does not excite fluxes but can heat the matter sector via the quartic interacting Hamiltonian of itinerant Majorana fermions. We note that our driving scheme is chosen for a crisp illustration of {\it fractional prethermalization}, e.g., allowing numerical feasibility. However, its behavior is generic as it relies on the basic observation that driving the physical spin degrees of freedom couples in general asymmetrically to the intrinsic fractionalized excitations.   
	
	Starting with the ground state $|\Psi_{0}\rangle$ as the initial state, the stroboscopic $t=NT$ time evolution is obtained from  $$U(NT)=\mathcal{T}_t\exp\left(-{\rm i}\int_0^{NT}H(t){\rm d}t\right)\equiv\exp\left(-{\rm i}NTH_{\rm eff}\right),$$ where $\mathcal{T}_t$ ensures that the exponential is time-ordered. Using the  Magnus expansion~\cite{Goldman2014, Bukov2015, Eckardt2017}, the effective Hamiltonian up to the first order reads $H_{\rm eff} = H_{\rm K}-\frac{i\pi}{2\omega}[H_{\rm K},H_{\rm V}+H_{\rm J}]+\mathcal{O}(\omega^{-2})$. In high-frequency limit, we thus recover the Kitaev honeycomb model $H_{\rm K}$ as the effective Hamiltonian that describes the prethermal regime.
	
	\textbf{{\em Fractionalized Prethermalization.---}}The different dynamics for gauge and matter sectors can be diagnosed by constructing suitable observables. The thermalization of static flux excitations can be captured by the dynamics of plaquette operators, $W(t)\equiv\sum_{p}\langle{}W_{p}\rangle_t$, where the expectation value $\langle{}\cdot\rangle_t$ is obtained with respect to the time-evolved state $|\Psi_{0}(t)\rangle = U(t) |\Psi_{0}\rangle$. We moreover keep track of energy absorption by measuring the energy of the effective Hamiltonian $E(t)\equiv\langle{}H_{\rm K}\rangle_t$. Even though the excitations of both fractionalized particles can contribute to the total energy $E(t)$, it can act as a measure for the thermalization of the itinerant Majorana fermions in the regime in which the fluxes are almost frozen.
	We compute both observables at stroboscopic times $t=NT$. The time evolution and stroboscopic measurements are numerically implemented with exact-diagonalization on a $4\times{}3$ torus with $24$ spins (qubits). We  explicitly impose translational symmetries along both directions of the torus and work in the zero momentum sector.

	In accordance with the prethermalization paradigm, the system can get stuck in a prethermal regime for an exponentially long time $\sim{}e^{c\omega}$ when the drive frequency $\omega$ exceeds a critical value. 
	As shown in Fig.~\ref{fig:fig1}(b) and (c), we find that the driven Kitaev spin liquid can exhibit different prethermalization behaviors: (i) When $\omega<\omega_{1}$, both the energy $E(t)$ and flux $W(t)$ quickly decay to zero and a conventional steady-state is reached in which both flux and matter sectors are at infinite temperature. The flux sector, in particular, is activated by the Heisenberg interaction as can be confirmed by rescaling time with $J^2$; see supplemental material~\cite{appendix}. 
	(ii) For intermediate drive frequencies $\omega_{1}<\omega<\omega_{2}$, the system enters a prethermal regime in which the flux $W(t)$ remains close to the ground state value for an exponentially long time $\sim{}e^{c\omega}$. At the same time, the energy $E(t)$ is already fluctuating around a small value close to zero corresponding to a high-temperature state. The freezing of fluxes $W(t)$ signals that in this regime the excitations of thermally-activated itinerant Majorana fermions mostly contribute to the energy growth. The prethermal regime thus cannot be described by a conventional thermal Gibbs state of an effective Hamiltonian. Rather, the fractionalized matter and flux degrees of freedom are at two distinct temperatures. (iii) For high drive frequency $\omega>\omega_{2}$, not only the flux remains in its ground state, but also the energy absorption of matter fermions is inefficient leading to prethermal plateaus in both quantities.

	\begin{figure}
		\centering
		\includegraphics[width=1\linewidth]{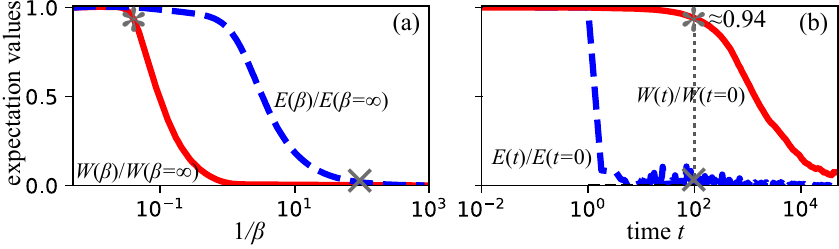}
		\caption{\textbf{Nonthermal heating.} (a) The equilibrium thermal expectation values of normalized energy (dashed) and flux (solid) as functions of temperature $1/\beta$. (b) The dynamics of $E(t)/E(t=0)$ (dashed) and $W(t)/W(t=0)$ (solid) in the Kitaev honeycomb model driven at frequency $\omega=8$. The model parameters are the same as those in Fig.~\ref{fig:fig1}. At intermediate times $t = 100$, vertical grey line, we obtain for the fluxes $W(t)/W(t=0) \approx 0.94$ (star) and for the energy $E(t)/E(t=0) \approx 0.03$ (cross), which are effectively at temperatures $1/\beta_\text{flux} \approx 0.04$ and $1/\beta_\text{matter} \approx 50$ when comparing with the thermal expectation values as indicated by star and cross in (a).}
		\label{fig:fig2}
	\end{figure}
	
	Next, we investigate signatures of fractionalized prethermalization in the dynamics of the entanglement entropy $S(t)$ of the time evolved state $|\Psi_{0}(t)\rangle$. Dividing the torus into two equal sub-cylinders, we focus on the half-chain entanglement entropy between the two. One can observe two plateaus, showing a staircase-like heating process, see Fig.~\ref{fig:fig1}(d). The first plateau 
	in $S(t)$ corresponds to the thermalization of itinerant Majorana fermions in the matter sector before also the flux sector explores the full configurational space at much later times.

	It turns out that the entanglement entropy of the initial state $|\Psi_{0}\rangle$ can be expressed in a separable form $S_0=S_{\rm G}+S_{\rm M}$, where $S_{\rm G}$ and $S_{\rm M}$ are the entropy of $Z_2$ gauge fields and itinerant Majorana fermions, respectively~\cite{Yao2010}. In order to quantify the above numerical results, we obtain the entanglement of an infinite-temperature state in the matter sector, by computing the entanglement of a random vector in the Hilbert space of itinerant Majorana fermions only. This entanglement corresponds to the Page saturation value of matter fermions, taking into account all the non-trivial conservation laws. The difference between the entanglement of the infinite-temperature state and the ground state entanglement $S_{\rm M}$ in the matter sector is $\approx0.21\mathrm{N}_{\rm c}$ ($\mathrm{N}_{\rm c}=12$ is the size of sub-cylinder), which is consistent with the entropy increase found from the time evolved state in Fig.~\ref{fig:fig1}(d). After an exponentially long time, $S(t)$ reaches $S(\infty)\approx{}0.66\mathrm{N}_{\rm c}$ which indeed corresponds to the entanglement of a fully random state covering both the matter and flux sectors. Thus a true infinite temperature state is reached. Note, the entropy per site deviates slightly from the maximum possible value of log(2) due to the finite-size corrections and the imposed translational symmetries.
	
	We have shown that our periodic drive thermalizes the matter sector more efficiently than the flux sector leading to a novel staircase prethermalization profile of the entanglement entropy. This is in stark contrast compared to a conventional thermal equilibrium state. In thermal equilibrium, fluxes are excited at a finite density as determined by their finite-temperature Boltzmann weight. In order to quantitatively analyze this difference, we compute the thermodynamic expectation values of the (normalized) energy $E_{\beta}$ and flux $W_{\beta}$ as a function of temperature [Fig.~\ref{fig:fig2} (a)]; see supplemental material for details~\cite{appendix}. For the Kitaev honeycomb model prepared in an equilibrium state at intermediate temperature $\beta \approx 1$, 
	the fluxes are already thermally activated $W\approx{}0$, while the corresponding energy is still close to the zero-temperature value, $E\approx0.9E(\beta=\infty)$~\cite{nasu2015thermal}. Thus, in equilibrium the flux sector is much stronger affected than the matter sector as the temperature increases. 
	
	Without fractionalized excitations, one could expect generic Floquet prethermalization paradigm, in which the system is governed by the effective Hamiltonian $H_{\rm K}$ at an effective temperature that is set by the energy pumped into the system. By contrast, the dynamical Hamiltonian in Eq.~\eqref{eq:Ht} exhibits a completely different prethermalization behavior. At intermediate drive frequencies, the effective time to thermalize flux sector takes two orders of magnitude longer than that for itinerant Majorana fermions. Specifically, the energy $E(t)$ quickly drops to (almost) zero at time $t \approx 1$, while the flux $W(t)$ still remains frozen for times $t \approx 100$. 
	When inspecting, for example, times $t \approx 100$, which are well within the fractionalized prethermal plateau for $\omega = 8$, the temperatures corresponding to the expectation values of the matter and flux sector are $1/\beta_\text{matter} \approx 50$ and $1/\beta_\text{flux} \approx 0.04$, respectively. Hence they differ by about three orders of magnitude. While we argue that the phenomenon is generic as any periodic modulation of physical spins typically couples non-symmetrically to the fractionalized excitations, the unusual large separation of heating times arises in our Floquet protocol because the Heisenberg interaction is small compared to the three-spin term defined in Eq.~\eqref{eq:HV}, and the latter only heats the matter sector. 
	
	\begin{figure}
		\centering
		\includegraphics[width=\linewidth]{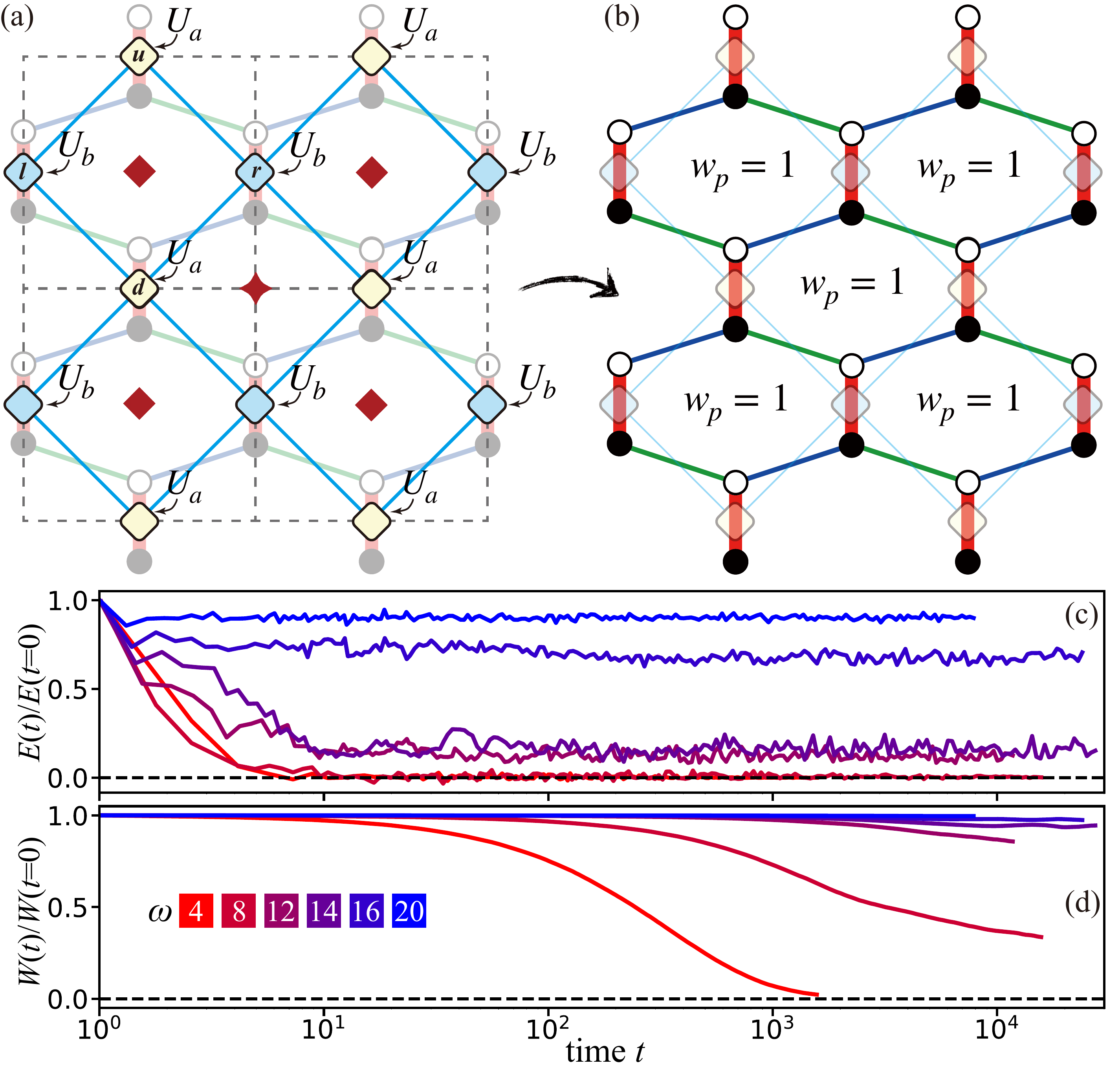}
		\caption{\textbf{Preparation of a zero-flux state $|\tilde{{\Psi}}_0\rangle$.} (a) Representation of a toric code state of effective spins $\bf{\tau}$ on the super lattice (dashed square) formed by the $z$-bonds of the Kitaev honeycomb model.  The blue (yellow) diamonds indicate the horizontal (vertical) sublattices on which the effective spins $\bf{\tau}$ live.  (b) After applying a unitary transformation $U_{ab}$ on the toric code state, we obtain a zero-flux state $|\tilde{{\Psi}}_{0}\rangle$. (c, d) The system in Eq.~\eqref{eq:Ht} initialized with $|\tilde{{\Psi}}_0\rangle$ exhibits similar nonequilibrium fractionalized dynamics as obtained with the ground state of Kitaev honeycomb model $|{\Psi}_0\rangle$. The simulations are performed for $V=1.0$ and $J=0.02$. }
		\label{fig:zero-flux}
	\end{figure}
	
	{\textbf{ \em Experimental feasibility.}---}One intriguing prospect is to experimentally observe fractionalized prethermalization. First, we discuss the implementation of the dynamical Hamiltonian, which consists of three terms, the Kitaev honeycomb model $H_{\rm K}$, the Heisenberg interaction $H_{\rm J}$, and the three spin interaction $H_{\rm V}$. The first two can be directly decomposed into two-qubit Ising gates which in principle can be realized in various quantum architectures such as superconducting quantum processor~\cite{You2010,Sameti2019} and trapped atoms or molecules~\cite{Duan2003,Micheli2006, Bluvstein2022}. Moreover, the three spin term can also be conveniently prepared with two-qubit Ising gates by noting that $\sigma^x_j\sigma^y_k\sigma^z_l\propto(\sigma^x_j\sigma^x_i)(\sigma^y_k\sigma^y_i)(\sigma^z_l\sigma^z_i)$.
	Second, the experimental preparation of $|\Psi_0\rangle$, the ground state of Kitaev honeycomb model, as an initial state is highly nontrivial. However, we need not to start in the ground state of the model, but a flux eigenstate at low energy density is sufficient. Thus, we propose a zero-flux state $|\tilde{{\Psi}}_{0}\rangle$ which can be more easily realized in experiments and can lead to similar results as those obtained with $|\Psi_{0}\rangle$.
	Our proposal for preparing $|\tilde{{\Psi}}_{0}\rangle$ is motivated  by the idea that the ground state of the Kitaev honeycomb model in the gapped $A$ phase is continuously connected to a toric code state~\cite{Kitaev06}. We can, therefore, leverage previous work for the preparation of $|\tilde{{\Psi}}_{0}\rangle$, which showed that a toric code state can be efficiently prepared with a finite-depth quantum circuit~\cite{Satzinger2021}. 
	
	On the honeycomb lattice all $z$-type bonds form a super lattice, i.e., a square lattice shown in Fig.~\ref{fig:zero-flux}(a).  We introduce an effective spin $\bm{\tau}=(\tau^x,\tau^y,\tau^z)$, that lies on each link of the square lattice, with a new local basis of $(|\Uparrow\rangle\equiv|\uparrow\uparrow\rangle,|\Downarrow\rangle\equiv|\downarrow\downarrow\rangle)$. This basis indeed spans the ground-state manifold of the Kitaev honeycomb model with $K_x=K_y\rightarrow0$ and $K_z>0$. The plaquette operators $W_p$ can be rewritten in terms of effective spins $\bm{\tau}$ as	$W_p\rightarrow\tilde{W}_p=\tau^z_u\tau^y_l\tau^z_d\tau^y_r$, where the super-lattice sites $u$, $l$, $d$, and $r$ are shown in Fig.~\ref{fig:zero-flux}(a). We further divide this super-lattice into to two sublattices, the vertical and horizontal super-lattice sites marked by blue and yellow diamonds, respectively. The effective plaquette operators $\tilde{W}_p$ with site $u$ belonging to the horizontal (vertical) sublattice are defined on the plaquettes (vertices) of the super lattice. 
	
	
	We can now introduce a quantum circuit to prepare $|\tilde{\Psi}_0\rangle$ on a finite-size cluster by the following steps (here we use a cluster with 24 qubits as an example):
	(i) Begin with a product state $|\uparrow\rangle^{\otimes24}=|\Uparrow\rangle^{\otimes{}12}$.
	(ii) Prepare a toric code state in the basis of $|\Uparrow\rangle$ and $|\Downarrow\rangle$ by using the quantum circuit introduced in Ref.~\cite{Satzinger2021}.
	(iii) Apply a unitary transformation $U_{ab}=\prod_{\rm horizontal\ sites}U_a\prod_{\rm vertical\ sites}U_b$ to the toric code state, where $U_a\equiv{}e^{i\tau^y\pi/4}e^{i\tau^x\pi/4}$ and
	$U_b\equiv{}e^{-i\tau^z\pi/4}$~\cite{appendix}. This unitary transforms the star and plaquette operators of the toric code to plaquette operators of the Kitaev model, $Z_s=\tau^z_u\tau^z_l\tau^z_d\tau^z_r \to \tilde{W}_{p\in {\rm vertices}}$ and $X_p=\tau^x_u\tau^x_l\tau^x_d\tau^x_r \to \tilde{W}_{p\in{\rm plaquette }}$ and thus yields $|\tilde{\Psi}_{0}\rangle$  with all $w_p=1$. Note that two-qubit gates, such as $U_a$ and $U_b$, can be decomposed into several CNOT gates and one-qubit gates~\cite{Kraus2001,Vatan2004,Smith2019}. Crucially, the depth of the circuit is linear in system size, see supplement~\cite{appendix}. 
	
	For this zero-flux state $|\tilde{{\Psi}}_0\rangle$ we now evaluate the heating dynamics on a plane with 24 qubits under drive of Eq.~\eqref{eq:Ht}; Fig.~\ref{fig:fig1}(b). 
	The energy of $|\tilde{{\Psi}}_0\rangle$ is $0.57E_0$ with $E_0\approx-21.3K$ the ground-state energy of a Kitaev honeycomb model on such a lattice. Thus, the toric-code inspired state preparation leads to a finite effective temperature for the matter sector, while the fluxes remain at zero temperature. Generally, the detail of the dynamics depends on the excitations injected in the initial state. Remarkably, we observe a multi-stage relaxation dynamics also for the initial state $|\tilde{{\Psi}}_0\rangle$, and hence fractionalized prethermalization is robust phenomenon; see Fig.~\ref{fig:zero-flux}(c) and (d). 
	
	{\textbf{\em Discussion.---}}We have shown that Floquet driven systems can exhibit unusual heating behavior in the presence of fractionalized excitations. Despite driving a physical degree of freedom of an ergodic system, we establish the emergence of distinct prethermal plateaus characterized by different temperatures for fractionalized excitations. 
	Our concrete example of the driven Kitaev honeycomb model confirmed that in the fractional prethermal regime the matter and flux sectors are governed by two different temper   atures. 
	
	In contrast to the thermal equilibrium states of the Kitaev honeycomb model, the matter sector thermalizes more efficiently than the flux sector in our driving protocol because of the three-spin interaction $H_{\rm V}$. Though $H_{\rm V}$ can perturbatively emerge in the presence of a magnetic field, we found that driving a magnetic field term unavoidably heats up the flux sector first. Nevertheless, the concept of fractionalized prethermalization is generic and insensitive to the details of the drive protocol because generically a drive couples asymmetrically to fractionalized excitations. 
	
	For future work it will be very worthwhile to study other examples of driven fractionalized quantum many-body phases, e.g., the one-dimensional Hubbard model with spin-charge separation, as they might show similarly rich fractionalized prethermalization physics. Moreover, it will be interesting to study whether classical spin liquids can exhibit a related phenomenology. Often quantum spin liquids emerge in the low-energy manifold of certain systems. Investigating under which conditions fractional prethermalization occurs in such systems will be pertinent as well. The Kitaev spin liquid studied here exhibits a block diagonal Hilbert space structure on all energy scales~\cite{baskaran2007exact}. Hence, weak perturbations are expected to similarly affect the whole spectrum, which renders the fractionalized prethermalization phenomena of the Kitaev spin liquid rather robust. 
	
	We emphasize that while fractionalization generically leads to two-temperature prethermal states as we discuss, the converse is not true; see e.g. Ref.~\cite{Lindner2017}. 
	Since the experimental identification of quantum spin liquids is notoriously difficult~\cite{knolle2019field}, an exciting possibility would be to use fractionalized prethermalization as a signature of fractionalization. 
	In conclusion, our work considerably enriches the phenomenology of driven phases of matter and we expect that fractionalized phases will be a versatile area for exotic non-equilibrium physics.
	
	\textbf{{\em Note added.---}}While finalizing this manuscript Ref.~\cite{Kalinowski2022} appeared, which proposed the same toric-code based protocol for preparing a zero-flux state in the Kitaev honeycomb model. 
	
	\textbf{{\em Data and materials availability.---}}Data analysis and simulation codes are available on Zenodo upon reasonable request~\cite{zenodo}.
	
	\begin{acknowledgments}
		\textbf{{\em Acknowledgments.---}}We thank Roderich Moessner, Andrea Pizzi, and Hongzheng Zhao for helpful discussions.
		The numerical simulations in this work are based on the Quspin project~\cite{quspin1}. We acknowledge support from the Imperial-TUM flagship partnership, the Deutsche Forschungsgemeinschaft (DFG, German Research Foundation) under Germany’s Excellence Strategy--EXC--2111--390814868 and DFG grants No. KN1254/1-2, KN1254/2-1, the European Research Council (ERC) under the European Union’s Horizon 2020 research and innovation programme (Grant Agreements No. 771537 and No. 851161), as well as the Munich Quantum Valley, which is supported by the Bavarian state government with funds from the Hightech Agenda Bayern Plus.
	\end{acknowledgments}

	\bibliography{heatKitaev.bib}

	\clearpage
	\newpage
	
	\begin{widetext}
		\centering
		{\large \textbf{ Supplemental Materials for \\``Fractionalized Prethermalization in a Driven Quantum Spin Liquid''}}
		
		\
		\
		
		{Hui-Ke Jin$^1$}, {Johannes Knolle$^{1,2,3}$}, {Michael Knap$^{1,2}$}
		
		{\em 
			{$^1$Department of Physics, Technical University of Munich, 85748 Garching, Germany}
			
			{$^2$Munich Center for Quantum Science and Technology (MCQST), Schellingstr. 4, 80799 Munich, Germany}
			
			{$^3$Blackett Laboratory, Imperial College London, London SW7 2AZ, United Kingdom}
		}
	\end{widetext}

	\setcounter{equation}{0}
	\setcounter{figure}{0}
	\renewcommand{\theequation}{S\arabic{equation}}
	\renewcommand{\thefigure}{S\arabic{figure}}
	\renewcommand{\thetable}{S\arabic{table}}
	
	In this Supplemental Material, we provide additional information and details for some of the technical aspects of our work. 
	
	\section{Thermal expectation value of energy and flux}
	We recall the four Majorana representation for $S=1/2$ spins at site $j$:
	\begin{eqnarray}
		\sigma^a_j=ic^a_jc^0_j.
	\end{eqnarray}
	The Hamiltonian $H_{\rm K}$ can be written in terms of Majoranas:
	\begin{equation}
		H_{\rm K}=-iK\sum_{\langle{}ij\rangle_a}u_{ij}c^0_ic^0_j,
	\end{equation}
	where $u_{ij}=ic^a_ic^a_j$. We follow a convention that for one $u_{ij}$, $i$ belongs to sublattice $A$ and $j$ belongs to sublattice $B$. 
	A given set of $\{u_{ij}\}$ can reduce $H_{\rm K}$ to a quadratic form in $c^0$:
	\begin{equation}
		H_{\rm K}(\{u_{ij}\})=\frac{i}{2}\left((c^0_A)^T, (c^0_B)^T\right)\left(\begin{array}{cc}
			0 & M \\
			-M^T & 0
		\end{array}\right)\left(\begin{array}{c}
			c^0_A \\
			c^0_B
		\end{array}\right)\label{eq:HKmat},
	\end{equation}
	where $c^0_{A(B)}$ is a column vector of itinerant Majoranas on the $A(B)$ sublattice and $M_{ij}=Ku_{ij}$ if site $i$ on $A$ sublattice and $j$ on $B$ sublattice are connected by the nearest neighbor bonds of a honeycomb lattice and $M_{ij}=0$ otherwise.
	
	Hamiltonian in Eq.~\eqref{eq:HKmat} can be diagonalized by singular-value decomposition (SVD) of the $N\times{}N$ matrix $M$ ($N$ the number of unit cell of a honeycomb lattice), $M=U\Lambda{}V^T$. Here $U^TU=I$, $V^TV=I$, and $\Lambda{}_{mn}=\delta_{nm}\epsilon_m$ is a semi-positive diagonal matrix. Then we can define new Majorana fermions as 
	\begin{equation}
		\begin{split}
			&(\alpha_1, ..., \alpha_N) = (c^0_{A,1}, ..., c^0_{A,N}) U,\\
			&(\beta_1, ..., \beta_N) = (c^0_{B,1}, ..., c^0_{B,N}) V.\\
		\end{split}
	\end{equation}
	And $H_{\rm K}(\{u_{ij}\})$ reads 
	\begin{eqnarray}
		H_{\rm K}(\{u_{ij}\}) = i\sum_{m}\epsilon_m\alpha_m\beta_m=\sum_{m}\epsilon_m\left(2a^\dagger_ma_m-1\right),\label{eq:Hku}
	\end{eqnarray}
	where $a_m=\frac{1}{2}(\alpha_m+i\beta_m).$ Because $\epsilon_m\geq{}0$ for all $m$, the ground state of $H_{\rm K}(\{u_{ij}\})$ is a vacuum of $a_m$ fermions with ground-state energy $-\sum_m\epsilon_m$. And the excitations of $a$-fermions generates the whole spectrum of $H_{\rm K}$ for a given $Z_2$ gauge configuartion.
	
	By iterating over all possible flux configurations, the whole spectrum of $H_{\rm K}$ in the enlarged Majorana Hilbert space can be obtained according to Eq.~\eqref{eq:Hku}. However, half of the eigenstates in the enlarged Hilbert are unphysical. When calculating the thermal expectation values, we should only account for the contribution from physical states.
	
	\begin{figure*}[!htp]
		\includegraphics[width=\linewidth]{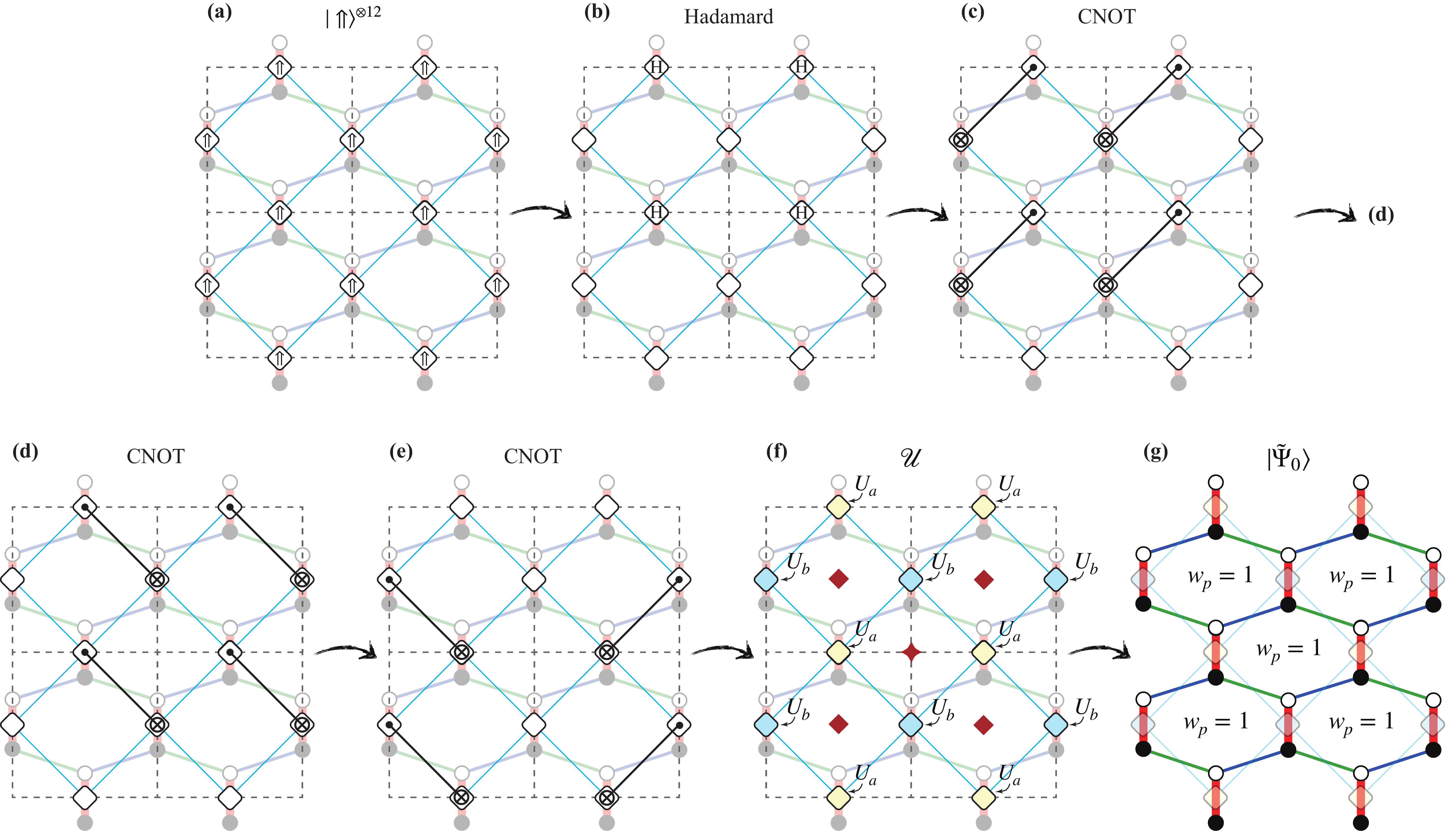}
		\caption{\textbf{Preparation of a zero-flux state $|\tilde{{\Psi}}_0\rangle$.} (a) Initialize the system with a product state $|\uparrow\rangle^{\otimes{}24}=|\Uparrow\rangle^{\otimes{}12}$. (b) Apply Hadamard gates $H$ on all representative effective spins. (c)-(e) Apply CNOT gates to obtain a toric code state in terms of effective spins. (f) Apply a unitary transformation ${U}_{ab}$, which consists of unitaries $U_a$ and $U_b$ applied on the horizontal and vertical sites, respectively. (g) This algorithm realizes a zero-flux state for the Kitaev honeycomb model on a surface.}\label{fig:sm1}
	\end{figure*}
	
	A physical eigenstate of $H_{\rm K}(\{u_{ij}\})$, $|\psi_n\rangle$, satisfies the condition $D_j|\psi_n\rangle=|\psi_n\rangle$ for all lattice sites $j$ with $D_j\equiv c^{x}_jc^{y}_jc^{z}_jc^0_j$. It means that the physical subspace of the Majorana Hilbert space can be determined by a projection $\mathcal{P}$
	\begin{eqnarray}
		\mathcal{P}=\prod_{j}\left(\frac{1+D_j}{2}\right).
	\end{eqnarray}
	The effect of $\mathcal{P}$ is to annihilate unphysical states. In accordance with Ref.~\cite{Yao2009}, $\mathcal{P}$ can be alternatively written as
	\begin{equation}
		\mathcal{P}=\mathcal{S}\left(\frac{1+\mathcal{D}}{2}\right),
	\end{equation}
	where $\mathcal{S}$ symmetrizes over all gauge-equivalent flux configurations and $\mathcal{D}=\prod_jD_j$. Introducing a $2N\times{}2N$ matrix
	\begin{eqnarray}
		Q=\left(\begin{array}{cc}
			0 & U\\
			V & 0 
		\end{array}\right),
	\end{eqnarray}
	the operator $\mathcal{D}$ reads~\cite{Pedrocchi2011,Zschocke2015}
	\begin{equation}
		\mathcal{D}=(-1)^\theta{\rm det}(Q)(-1)^{N_a}\prod_{\langle{}ij\rangle_a}u_{ij},~\label{eq:D}
	\end{equation}
	where $N_a=\sum_{m}a^\dagger_ma_m$. In our case, e.g., a torus without any twisted boundaries, $\theta=L_x+L_y$ with $L_x$ and $L_y$ the length of a torus along $x$ and $y$ directions, respectively. By employing Eq.~\eqref{eq:D}, all of the physical eigenstates can be selected.
	
	The partition function for $H_{\rm K}$ at temperature $1/\beta$ is 
	\begin{eqnarray}
		Z(\beta)=\sum_{\{w_p\}}\sum_{E_n}\langle{}\psi_n(\{u_{ij}\})|\mathcal{P}e^{-\beta{}E_n}\mathcal{P}|\psi_n(\{u_{ij}\})\rangle
	\end{eqnarray}
	where $\sum_{\{w_p\}}$ sums over all possible flux configurations, $\{u_{ij}\}$ is a $Z_2$ gauge field configuration leading to the given flux configuration $\{w_p\}$, and $\sum_{E_n}$ sums over all eigenstates with eigenenergies $E_n$ for given $\{u_{ij}\}$.
	The thermal expectation value of energy $E(\beta)$ and flux $W(\beta)$ reads
	\begin{equation}
		\begin{split}
			&E(\beta)=\sum_{\{w_p\}}\sum_{E_n}\langle{}\psi_n(\{u_{ij}\})|\mathcal{P}E_ne^{-\beta{}E_n}\mathcal{P}|\psi_n(\{u_{ij}\})\rangle/Z(\beta),\\
			&W(\beta)=\sum_{\{w_p\}}\sum_{E_n}\left\langle{}\psi_n(\{u_{ij}\})\left|\mathcal{P}{}\sum_{p}w_pe^{-\beta{}E_n}\mathcal{P}\right|\psi_n(\{u_{ij}\})\right\rangle/Z(\beta).
		\end{split}		
	\end{equation}

	\section{Details for preparing a zero-flux state $|\tilde{{\Psi}}_0\rangle$}
	In this section, we provide a finite-depth quantum circuit  for preparing $|\tilde{{\Psi}}_0\rangle$, a zero-flux state for the Kitaev honeycomb model on a finite size lattice. As mentioned in the main text, this zero-flux state is continuously connected to a toric code state. 
	
	For completeness, we first review the general circuit design principle for the toric code state introduced in Ref.~\cite{Satzinger2021}. This quantum circuit can be divided into the following main steps: (i) Initialize all qubits in a simple product state $|\uparrow\uparrow\uparrow...\uparrow\rangle$, see Fig.~\ref{fig:sm1}(a). (ii) Find the optimal representative qubits for a given geometry, and then apply a Hadamard gate $H$ on each representative qubit, see Fig.~\ref{fig:sm1}(b). (iii)  Perform the CNOT gates around each plaquette in a specific order so that the states in representative qubits are not changed until all the CNOT gates in their plaquette have been applied, see Fig.~\ref{fig:sm1}(c)-(e).

	Notice that the toric code state is prepared in terms of effective spins $\tau$ spanned by the two-qubit state $|\uparrow\rangle\rightarrow{}|\uparrow\uparrow\rangle\equiv{}|\Uparrow\rangle$ and $|\downarrow\rangle\rightarrow{}|\downarrow\downarrow\rangle\equiv{}|\Downarrow\rangle$.  Here we explicitly write down the effective spins in the basis of $(|\uparrow\uparrow\rangle\equiv|\Uparrow\rangle,~|\uparrow\downarrow\rangle,~|\downarrow\uparrow\rangle,~|\downarrow\downarrow\rangle \equiv|\Downarrow\rangle)$ as
	\begin{equation*}
		\begin{split}
			&\tau^x=\left(\begin{array}{cccc}
				0 & & & 1  \\
				& 0 & &  \\
				&  & 0 &  \\
				1 & & & 0  \\
			\end{array}\right),	
			\tau^y=\left(\begin{array}{cccc}
				0 & & & -i  \\
				& 0 & &  \\
				&  & 0 &  \\
				i & & & 0  \\
			\end{array}\right),	
			\tau^z=\left(\begin{array}{cccc}
				1 & & &   \\
				& 0 & &  \\
				&  & 0 &  \\
				& & & -1  \\
			\end{array}\right).
		\end{split}
	\end{equation*}
	Then, the Hadamard gate  for the effective spins reads
	\begin{equation}
		H=\left(\begin{array}{cccc}
			1/{\sqrt{2}} & & &  1/{\sqrt{2}} \\
			& 1 & &  \\
			&  & 1 &  \\
			1/{\sqrt{2}} & & & -1/{\sqrt{2}} \\
		\end{array}\right),
	\end{equation}
	Actually, the $2\times2$ identity submatrix inside the Hadamard gate can be an arbitrary $2\times2$ unitary matrix since this submatrix only acts on a null space. Similarly, the CNOT gate in terms of effective spins reads
	\begin{eqnarray}
		\begin{split}
			\mbox{CNOT}  = &~|\Uparrow_c\Uparrow_t\rangle\langle\Uparrow_c\Uparrow_t| + |\Uparrow_c\Downarrow_t\rangle\langle\Uparrow_c\Downarrow_t| +\\
			&~|\Downarrow_c\Uparrow_t\rangle\langle\Downarrow_c\Downarrow_t| + |\Downarrow_c\Downarrow_t\rangle\langle\Downarrow_c\Uparrow_t|	+ U_{\rm arb },
		\end{split}~\label{eq:CNOT}
	\end{eqnarray}
	where the subindex $c$ ($t$) stands for the control (target) two-site qubit and again $U_{\rm arb}$ is an arbitrary $12\times{}12$ unitary matrix spanned by $(|\uparrow_c\downarrow_c\rangle,|\downarrow_c\uparrow_c\rangle)\otimes{}(|\Uparrow_t\rangle,|\uparrow_t\downarrow_t\rangle,|\downarrow_t\uparrow_t\rangle,|\Downarrow_t\rangle)\oplus{}(|\Uparrow_c\rangle,|\Downarrow_c\rangle)\otimes(|\uparrow_t\downarrow_t\rangle,|\downarrow_t\uparrow_t\rangle)$. Note that the CNOT gate in Eq.~\eqref{eq:CNOT} involves four physical qubits.

	\begin{figure*}[!ht]
		\includegraphics[width=0.95\linewidth]{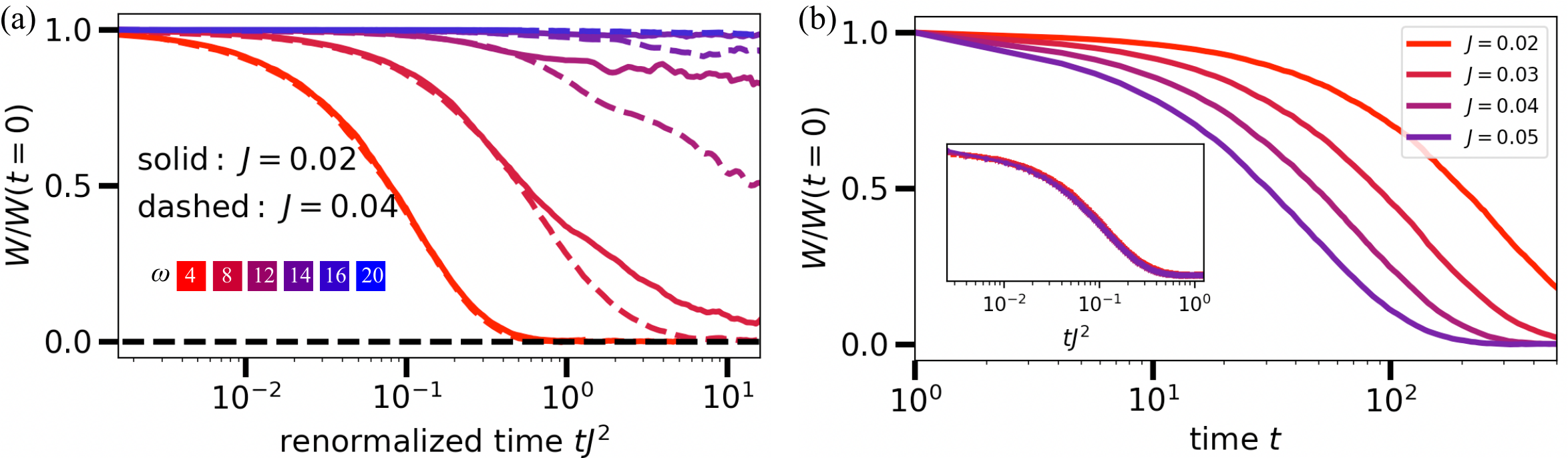}
		\caption{{\bf Mechanism for the heating of the flux degree of freedom.} (a) The dynamics of renormalized flux on short time scales for various drive frequencies $\omega$. (b) The same observable but with fixed $\omega=4$. The inset demonstrates a collapse of the short time flux dynamics $W(t)$ when times are rescaled by $J^2$. The simulations are performed for $V=1.0$ on a $3\times{}4$ torus.}\label{fig:collapse}
	\end{figure*}
	
	After obtaining the state in Fig.~\ref{fig:sm1}(e), one can apply a unitary transformation $U_{ab}$ to the toric code state to obtain $|\tilde{\Psi}_{0}\rangle$, a zero-flux stare for Kitaev honeycomb model, as shown in Fig.~\ref{fig:sm1}(f)-(g). 
	The explicit form of $U_a$ and $U_b$ in the basis of $\{|\uparrow\uparrow\rangle,~|\uparrow\downarrow\rangle,~|\downarrow\uparrow\rangle,~|\downarrow\downarrow\rangle\}$ is
	\begin{subequations}\label{eq:UvUh}
		\begin{equation}
			U_a=\left(\begin{array}{cccc}
				(1+i)/2 & & & (1+i)/2  \\
				& 1 & &  \\
				&  & 1 &  \\
				(-1+i)/2 & & & (1-i)/2 \\
			\end{array}\right),
		\end{equation}
		and
		\begin{equation}
			U_b=\left(\begin{array}{cccc}
				e^{-i\pi/4} & & & 0  \\
				& 1 & &  \\
				&  & 1 &  \\
				& & & e^{+i\pi/4}  \\
			\end{array}\right).
		\end{equation}
	\end{subequations}
	Similar to the Hadmard gates, the $2\times2$ identity submatrices in Eq.~\eqref{eq:UvUh} can be arbitrary $2\times2$ unitary matrices.
	
	\section{Mechanism of short-time flux heating }
	In order to examine which mechanism heats the flux sector, we vary the Heisenberg coupling $J$ and study the dynamics of $W(t)$. When rescaling the time with $J^2$ as $t\rightarrow{}tJ^2$, the flux dynamics $W(t)$ give a collapse of the short time data for various $J$, see Fig.~\ref{fig:collapse}. This data collapse can be explained from the following arguments. Denoting the initial state (a zero-flux state) as $|\psi_0\rangle$, the short-time dynamics of plaquette operator $W_p$ up to the second order of $J$ reads 
	\begin{equation}
		W(t)\sim\langle\psi_0 |(1+iTJ \tilde H_{\rm J})W_p (1-iTJ\tilde H_{\rm J})|\psi_0\rangle,
	\end{equation}
	where $\tilde H_{\rm J}=\sum_{\langle{}jk\rangle}{\bm \sigma}_j\cdot{}{\bm \sigma}_k$, i.e., we have singled out the coupling $J$ from the Heisenberg interaction.
	All the terms that are first order in $J$ vanish because at short times one can approximately consider that the state still has zero flux. Then to leading order the decay rate is
	\begin{equation}
		\frac{{\rm d}W(t)}{{\rm d}t}\sim{}J^2\langle\psi_0 |H_{\rm J}W_pH_{\rm J}|\psi_0\rangle.
	\end{equation}
	Upon rescaling time $t \to J^2t$ the decay is independent of microscopic parameters. We can thus conclude that the slow heating of fluxes is due to the weak coupling to the periodic drive.

\end{document}